\documentstyle[11pt,amssymb]{article}

\begin{document}

\begin{flushleft}
\Large{\bf An explicit formula of the normalized Mumford form} 
\end{flushleft}

\begin{flushleft} 
\large{\bf Takashi Ichikawa} 
\end{flushleft}

\begin{flushleft} 
{\small Department of Mathematics, Faculty of Science and Engineering, Saga University, 
Saga 840-8502, Japan. E-mail: ichikawn@cc.saga-u.ac.jp} 
\end{flushleft} 

\noindent
{\bf Abstract:} \ 
We give an explicit formula of the normalized Mumford form which expresses 
the second tautological line bundle by the Hodge line bundle 
defined on the moduli space of algebraic curves of any genus. 
This formula is represented by an infinite product which is a higher genus version 
of the Ramanujan delta function under the trivialization 
by normalized abelian differentials and Eichler integrals of their products. 
By this formula, we have a universal expression of the normalized Mumford form 
as a computable power series with integral coefficients 
by the moduli parameters of algebraic curves. 
\vspace{3ex}

\vspace{3ex}

\noindent
{\bf 1. Introduction} 
\vspace{2ex} 

\noindent 
For integers $g > 1$ and $k > 0$, 
denote by ${\cal M}_{g}$ the moduli space of (algebraic) curves, 
and denote by $\lambda_{g,k}$ the $k$-th tautological line bundle on ${\cal M}_{g}$ 
whose each fiber is given by the determinant of the space of $k$-differentials 
(i.e., regular $k$-forms) on the corresponding curve. 
Then Mumford \cite{Mu2} that there exists an isomorphism 
$$
\mu_{g,k} : \lambda_{g,k} \stackrel{\sim}{\rightarrow} 
\lambda_{g,1}^{\otimes (6 k^{2} - 6 k + 1)}
$$ 
with certain boundary condition. 
Since ${\cal M}_{g}$ can be constructed as the moduli stack over ${\mathbb Z}$ 
(cf. \cite{DM}), 
$\mu_{g,k}$ is uniquely determined up to a sign, 
and then it is called the {\it normalized Mumford form}. 

In physics and mathematics, 
explicit formulas of $\mu_{g,k}$ are studied by Belavin-Knizhnik \cite{BelK}, 
Verlinde-Verlinde \cite{VV}, Beilinson-Manin \cite{BM}, Fay \cite{F}, 
Matone-Volpato \cite{MaV} and others, 
especially in the case $k = 2$ since $|\mu_{g,2}|^{2}$ is the Polyakov string measure 
(cf. \cite{BelK}). 
The formulas given in \cite{BM, F, MaV, VV} by theta functions 
use points on corresponding curves and do not express the normalized form. 
Then it is required to give a precise formula of the normalized Mumford form 
without using points on curves. 
When $g = 1$, $\mu_{1,2}$ is essentially the Ramanujan delta function 
$$
e^{2 \pi \sqrt{-1} z} \prod_{n=1}^{\infty} \left( 1 - e^{2 \pi \sqrt{-1} n z} \right)^{24} \ \ 
({\rm Im}(z) > 0), 
$$ 
and when $g = 2$ or $3$, 
$\mu_{g,2}$ becomes an integral Teichm\"{u}ller modular form of degree $g$ 
and is expressed in \cite{I2} by the product of even theta constants. 

The aim of this paper is to give a precise formula of 
the normalized Mumford form $\mu_{g,2}$ for any $g$ without using points on curves. 
Our formula is expressed by an infinite product which is a higher genus version 
of the Ramanujan delta function via the trivialization 
by normalized abelian (i.e., $1$-)differentials and Eichler integrals of their products. 
By this formula, one can obtain a universal expression of $\mu_{g,2}$ 
as a computable power series with integral coefficients 
by local coordinates on ${\cal M}_{g}$. 

This formula is obtained by combining 
the arithmetic Schottky uniformization theory  \cite{I1, I2} 
with the formulas of Zograf \cite{Z1, Z2} and of McIntyre-Takhtajan \cite{McT} 
on determinants of Laplacians on Riemann surfaces. 
A key point of the proof is show the fact that 
the exterior products of normalized $k$-differentials give 
a local generator of $\lambda_{g,k}$ in the case $k = 2$ 
by expanding Eichler integrals of certain products of normalized abelian differentials. 
This fact is extended in \cite{I3} for general $k$, 
and hence there exists a similar formula of any $\mu_{g,k}$ 
if one can take a basis of the space of $k$-differentials which consists of products of normalized abelian differentials 
(such products exist by a theorem of Max Noether). 
\vspace{3ex}


\noindent
{\bf 2. Normalized differential} 
\vspace{2ex} 

\noindent
{\it 2.1. Schottky uniformization.} 
A Schottky group $\Gamma$ of rank $g$ is a free group 
$\langle \gamma_{1},..., \gamma_{g} \rangle$ with generators 
$\gamma_{i} \in PSL_{2}({\mathbb C})$ which map Jordan curves  
$C_{i} \subset {\mathbb P}^{1}_{\mathbb C} = {\mathbb C} \cup \{ \infty \}$ 
to other Jordan curves $C_{-i} \subset {\mathbb P}^{1}_{\mathbb C}$ 
with orientation reversed, 
where the interiors of $C_{\pm 1},..., C_{\pm g}$ are mutually disjoint. 
Each element $\gamma \in \Gamma - \{ 1 \}$ is conjugated in 
$PSL_{2}({\mathbb C})$ to $z \mapsto q_{\gamma} z$ 
for some $q_{\gamma} \in {\mathbb C}^{\times}$ with $|q_{\gamma}| < 1$ 
which is called the {\it multiplier} of $\gamma$. 
Therefore, one has 
$$
\frac{\gamma(z) - a_{\gamma}}{\gamma(z) - b_{\gamma}} = 
q_{\gamma} \frac{z - a_{\gamma}}{z - b_{\gamma}} 
$$
for some element $a_{\gamma}, b_{\gamma}$ of ${\mathbb P}^{1}_{\mathbb C}$ 
called the {\it attractive}, {\it repulsive} fixed points of $\gamma$ 
respectively. 
Then the discontinuity set 
$\Omega_{\Gamma} \subset {\mathbb P}^{1}_{\mathbb C}$ 
under the action of $\Gamma$ has a fundamental domain $D_{\Gamma}$ 
which is given by the complement of the union of the interiors of $C_{\pm 1},..., C_{\pm g}$. 
The quotient space $\Omega_{\Gamma}/\Gamma$ is a (compact) Riemann surface 
of genus $g$ which we denote by $R_{\Gamma}$. 
Furthermore, by a result of Koebe, 
every Riemann surface of genus $g$ can be represented in this manner. 
A Schottky group $\Gamma$ is {\it marked} if its free generators 
$\gamma_{1},..., \gamma_{g}$ are fixed, 
and a marked Schottky group $(\Gamma; \gamma_{1},..., \gamma_{g})$ 
is {\it normalized} if 
$a_{\gamma_{1}} = 0$, $b_{\gamma_{1}} = \infty$ and $a_{\gamma_{2}} = 1$. 
By definition, the Schottky space ${\mathfrak S}_{g}$ of degree $g$ 
is the space of marked Schottky groups of rank $g$ 
modulo conjugation in $PSL_{2}({\mathbb C})$ which becomes 
the space of normalized Schottky groups of rank $g$ if $g > 1$. 
Then ${\mathfrak S}_{g}$ is a covering space of the moduli space of 
Riemann surfaces of genus $g$.

Let $R$ be a Riemann surface of genus $g > 1$, 
and $\left\{ \alpha_{1},..., \alpha_{g}, \beta_{1},..., \beta_{g} \right\}$ 
be a set of standard generators of $\pi_{1}(R, x_{0})$ for some $x_{0} \in R$ 
satisfying 
$$
\left( \alpha_{1} \beta_{1} \alpha_{1}^{-1} \beta_{1}^{-1} \right) \cdots 
\left( \alpha_{g} \beta_{g} \alpha_{g}^{-1} \beta_{g}^{-1} \right) = 1. 
$$ 
Then one can take a marked Schottky group 
$(\Gamma; \gamma_{1},..., \gamma_{g})$ such that $R = R_{\Gamma}$ 
and that each $C_{k}$ is homotopic to $\alpha_{k}$. 
Therefore, 
there is uniquely a basis $\varphi_{1},..., \varphi_{g}$ of holomorphic $1$-forms 
such that $\int_{\alpha_{j}} \varphi_{i}$ is equal to the Kronecker delta $\delta_{ij}$, 
and then the period matrix 
$\tau = \left( \int_{\beta_{j}} \varphi_{i} \right)$ becomes a symmetric matrix 
whose imaginary part is positive definite. 
For each Schottky group $\Gamma$ of rank $g$, 
$\{ \varphi_{1},..., \varphi_{g} \}$ is a basis of the space 
$H^{0} \left( R_{\Gamma}, \Omega_{R_{\Gamma}} \right)$ of 
holomorphic $1$-forms on $R_{\Gamma}$. 
Therefore, the {\it Hodge line bundle} $\lambda_{1}$ consisting of 
$\bigwedge^{g} H^{0} \left( R_{\Gamma}, \Omega_{R_{\Gamma}} \right)$ 
$(\Gamma \in {\mathfrak S}_{g})$ becomes a holomorphic line bundle 
on ${\mathfrak S}_{g}$ with a holomorphic canonical section 
$\varphi_{1} \wedge \cdots \wedge \varphi_{g}$. 
From the viewpoint of arithmetic geometry, 
we call $2 \pi \sqrt{-1} \varphi$ $(1 \leq i \leq g)$ the {\it normalized abelian differentials}. 
\vspace{2ex}   

\noindent
{\it 2.2. Eichler integral and normalized differential.} 
Assume that $g > 1$, and take an integer $k > 1$. 
Let $\left( \Gamma; \gamma_{1},..., \gamma_{g} \right)$ be 
a marked normalized Schottky group, 
and ${\mathbb C}[z]_{2k-2}$ be the ${\mathbb C}$-vector space of polynomials 
$f = f(z)$ of $z$ with degree $\leq 2k-2$ on which $\Gamma$ acts as 
$$
\gamma(f)(z) = f \left( \gamma(z) \right) \cdot \gamma'(z)^{1-k} \ 
\left( \gamma \in \Gamma, \ f \in {\mathbb C}[z]_{2k-2} \right). 
$$
Take $\xi_{1,k-1}$, $\xi_{2,1},..., \xi_{2,2k-2}$, 
$\xi_{i,0},..., \xi_{i,2k-2}$ $(3 \leq i \leq g)$ 
as elements of the Eichler cohomology group 
$H^{1} \left( \Gamma, {\mathbb C}[z]_{2k-2} \right)$ 
of $\Gamma$ which are uniquely determined by the condition: 
$$
\xi_{i,j}(\gamma_{l}) = \left\{ \begin{array}{ll} 
\delta_{2l} (z-1)^{j} & (i = 2), 
\\ 
\delta_{il} z^{j} & (i \neq 2) 
\end{array} \right. 
$$
for $1 \leq l \leq g$. 
Then it is shown in \cite[Section 4]{McT} that the {\it Eichler integral} 
$$
\Psi_{g,k} (\psi, \xi) := 
\frac{1}{2 \pi \sqrt{-1}} \sum_{i=1}^{g} 
\oint_{C_{i}} \psi \cdot \xi(\gamma_{i}) dz 
$$ 
for $\psi (dz)^{k} \in H^{0} \left( R_{\Gamma}, \Omega^{k}_{R_{\Gamma}} \right)$, 
$\xi \in H^{1} \left( \Gamma, {\mathbb C}[z]_{2k-2} \right)$ 
is a non-degenerate pairing on 
$$
H^{0} \left( R_{\Gamma}, \Omega^{k}_{R_{\Gamma}} \right) \times 
H^{1} \left( \Gamma, {\mathbb C}[z]_{2k-2} \right). 
$$
Denote by 
$$ 
\left\{ \psi_{1,k-1}, \ \psi_{2,1},..., \psi_{2,2k-2}, \ \psi_{i,0},..., \psi_{i,2k-2} \ 
(3 \leq i \leq g) \right\} 
$$ 
the basis of $H^{0} \left( R_{\Gamma}, \Omega^{k}_{R_{\Gamma}} \right)$ 
dual to $\{ \xi_{l,m} \}$, 
namely $\Psi_{g,k} \left( \psi_{i,j}, \xi_{l,m} \right) = \delta_{il} \cdot \delta_{jm}$, 
and call $\psi_{i,j}$ {\it normalized $k$-differentials}. 
\vspace{2ex}

\noindent
{\it Remark.} 
Since $-\pi \cdot \Psi_{g,k}$ is the pairing given in \cite[4.1]{McT}, 
$$ 
\left\{ -\frac{\psi_{1,k-1}}{\pi}, \ -\frac{\psi_{2,1}}{\pi},..., -\frac{\psi_{2,2k-2}}{\pi}, \ 
-\frac{\psi_{i,0}}{\pi},..., -\frac{\psi_{i,2k-2}}{\pi} \ (3 \leq i \leq g) \right\} 
$$ 
is the natural basis for $k$-differentials defined in \cite{McT}. 
\vspace{2ex}

In what follows, put 
$$
\left\{ \psi^{(k)}_{1},..., \psi^{(k)}_{(2k-1)(g-1)} \right\} = 
\left\{ \psi_{1,k-1}, \ \psi_{2,1},..., \psi_{2,2k-2}, \ \psi_{i,0},..., \psi_{i,2k-2} \ 
(3 \leq i \leq g) \right\}, 
$$
and $\psi^{(k)} =  \psi^{(k)}_{1} \wedge \cdots \wedge \psi^{(k)}_{(2k-1)(g-1)}$. 
\vspace{3ex}

\noindent
{\bf 3. Explicit Mumford form} 
\vspace{2ex}

\noindent
{\it 3.1. Normalized Mumford form.} 
For an integer $g > 1$, 
let $\overline{\mathcal M}_{g}$ denote the moduli stack over ${\mathbb Z}$ 
of stable curves of genus $g$, 
and ${\mathcal M}_{g}$ denotes the open substack of $\overline{\mathcal M}_{g}$ 
classifying proper smooth curves of genus $g$ (cf. \cite{DM}). 
Denote by $\overline{\mathcal M}_{g}^{\rm an}$ and ${\mathcal M}_{g}^{\rm an}$ 
the complex orbifolds associated with 
$\overline{\mathcal M}_{g}$ and ${\mathcal M}_{g}$ respectively.  
By definition, there exists the universal curve 
$\pi : \overline{\mathcal C}_{g} \rightarrow \overline{\mathcal M}_{g}$. 
Then from the relative dualizing sheaf 
$\omega_{\overline{\mathcal C}_{g}/\overline{\mathcal M}_{g}}$ and the complement  $\partial {\mathcal M}_{g} = \overline{\mathcal M}_{g} - {\mathcal M}_{g}$ 
of ${\mathcal M}_{g}$, 
one can obtain the following line bundles over $\overline{\mathcal M}_{g}$: 
\begin{eqnarray*}
\lambda_{g,k} 
& : = & 
\det R \pi_{*} 
\left( \omega^{k}_{\overline{\mathcal C}_{g}/\overline{\mathcal M}_{g}} \right), 
\\ 
\delta_{g} 
& : = & 
{\mathcal O}_{\overline{\mathcal M}_{g}} \left( \partial {\mathcal M}_{g} \right). 
\end{eqnarray*} 
Furthermore, 
let $\kappa_{g}$ be the line bundle over $\overline{\mathcal M}_{g}$ defined as 
the following Deligne pairing: 
$$
\kappa_{g} = 
\left\langle \omega_{\overline{\mathcal C}_{g}/\overline{\mathcal M}_{g}}, \ 
\omega_{\overline{\mathcal C}_{g}/\overline{\mathcal M}_{g}} \right\rangle. 
$$
Then it is known (cf. \cite{D, Fr, GS, W}) that these line bundles have 
canonical hermitian metric over ${\mathcal M}_{g}^{\rm an}$ which is 
the Quillen metric \cite{Q} for $\lambda_{g,k}$. 
Furthermore, put 
$$
d_{k} = 6 k^{2} - 6 k + 1, \ \ 
a(g) = (2g - 2) \left( -12 \zeta'_{\mathbb Q}(-1) + \frac{1}{2} \right), 
$$
where $\zeta'_{\mathbb Q}(-1)$ denotes the derivative of the Riemann zeta function  $\zeta_{\mathbb Q}$ at $-1$. 
Then it is shown in \cite{D, Fr, GS, W} that there exists a unique (up to a sign) isomorphism 
$$
\rho_{g,k} : \lambda_{g,k}^{\otimes 12} \stackrel{\sim}{\rightarrow} 
\kappa_{g}^{\otimes d_{k}} \otimes \delta_{g} \cdot e^{a(g)} 
$$
between the line bundles over $\overline{\mathcal M}_{g}$ 
which is an isometry between the line bundles over ${\mathcal M}_{g}^{\rm an}$ 
for these hermitian structure. 
\vspace{2ex}

\noindent
{\bf Theorem 3.1.} 
\begin{it} 
There exists a unique (up to a sign) isomorphism 
$$
\mu_{g,k} : \lambda_{g,k} \stackrel{\sim}{\rightarrow} 
\lambda_{g,1}^{\otimes d_{k}} \otimes \delta_{g}^{\otimes (k - k^{2})/2} 
\cdot \exp \left( (k-k^{2}) a(g)/2 \right) 
$$
between the line bundles over $\overline{\mathcal M}_{g}$ 
which is also an isometry between the line bundles over ${\mathcal M}_{g}^{\rm an}$ 
for these hermitian structure. 
We call $\mu_{g,k}$ the Mumford isomorphism or the normalized Mumford form. 
\end{it} 
\vspace{2ex} 

\noindent
{\it Proof.} 
Mumford \cite{Mu2} shows the existence of $\mu_{g,k}$, 
and since $\rho_{g,k} = \rho_{g,1}^{\otimes d_{k}} \circ \mu_{g,k}^{\otimes 12}$, 
$\mu_{g,k}$ is an isometry. 
\ $\square$ 
\vspace{2ex} 

\noindent
{\it 3.2. Arithmetic Schottky uniformization.} 
The theory of arithmetic Schottky uniformization given in \cite{I2} 
constructs a higher genus version of the Tate curve, 
and its $1$-forms and periods. 
We review the theory for the special case concerned with 
universal deformations of irreducible degenerate curves. 

Denote by $\Delta$ the graph with one vertex and $g$ loops. 
Let $x_{\pm 1},..., x_{\pm g}$, $y_{1},..., y_{g}$ be variables, 
and put 
\begin{eqnarray*}
A_{g} & = & {\mathbb Z} \left[ x_{l}, \frac{1}{x_{m}-x_{n}} \ 
\left( l, m, n \in \{ \pm 1,..., \pm g \}, \ m \neq n \right) \right], 
\\
A_{\Delta} & = & A_{g} [[ y_{1},... y_{g} ]], 
\\
B_{\Delta} & = & A_{\Delta} \left[ 1/y_{i} \ (1 \leq i \leq g) \right]. 
\end{eqnarray*}
Then it is shown in \cite[Section 3]{I2} that 
there exists a stable curve $C_{\Delta}$ of genus $g$ over $A_{\Delta}$ 
which satisfies the followings: 
\begin{itemize}

\item[$\bullet$]  
$C_{\Delta}$ is a universal deformation of the universal degenerate curve 
with dual graph $\Delta$ which is obtained from ${\mathbb P}^{1}_{A_{g}}$ 
by identifying $x_{i}$ and $x_{-i}$ $(1 \leq i \leq g)$. 
The ideal of $A_{\Delta}$ generated by $y_{1},..., y_{g}$ corresponds to 
the boundary $\partial {\mathcal M}_{g} = \overline{\mathcal M}_{g} - {\mathcal M}_{g}$ 
of $\overline{\mathcal M}_{g}$ via the morphism 
${\rm Spec}(A_{\Delta}) \rightarrow \overline{\mathcal M}_{g}$ 
associated with $C_{\Delta}$. 

\item[$\bullet$]  
$C_{\Delta}$ is smooth over $B_{\Delta}$, 
and is Mumford uniformized (cf. \cite{Mu1}) by the subgroup $\Gamma_{\Delta}$ of 
$PGL_{2} \left( B_{\Delta} \right)$ 
with $g$ generators 
$$
\phi_{i} = 
\left( \begin{array}{cc} x_{i} & x_{-i} \\ 1 & 1 \end{array} \right) 
\left( \begin{array}{cc} y_{i} & 0      \\ 0 & 1 \end{array} \right) 
\left( \begin{array}{cc} x_{i} & x_{-i} \\ 1 & 1 \end{array} \right)^{-1} 
\ {\rm mod} \left( B_{\Delta}^{\times} \right) 
\ (1 \leq i \leq g). 
$$
Furthermore, 
$C_{\Delta}$ has the following universality: 
for a complete integrally closed noetherian local ring $R$ 
with quotient field $K$ 
and a Mumford curve $C$ over $K$ such that $\Delta$ is the dual graph of 
its degenerate reduction, 
there is a ring homomorphism $A_{\Delta} \rightarrow R$ 
which gives rise to $C_{\Delta} \otimes_{A_{\Delta}} K \cong C$. 

\item[$\bullet$]  
Let $\Gamma = \langle \gamma_{1},..., \gamma_{g} \rangle$ be a Schottky group of rank $g$, 
where each $\gamma_{i}$ has the attractive (resp. repulsive) fixed points $a_{i}$ (resp. $a_{-i}$), 
and it has the multiplier $q_{i}$. 
Then substituting $a_{\pm i}$ to $x_{\pm i}$ and $q_{i}$ to $y_{i}$ $(1 \leq i \leq g)$, 
$C_{\Delta}$ becomes the Riemann surface $R_{\Gamma}$ uniformized by $\Gamma$ 
if $|q_{i}|$ are sufficiently small. 

\end{itemize}

Actually, $C_{\Delta}$ is constructed in \cite{I2} as the quotient of 
a certain subspace of ${\mathbb P}^{1}_{B_{\Delta}}$ by the action of 
$\Gamma$ using the theory of formal schemes. 
Furthermore, as is shown in \cite{MD} and \cite[Section 3]{I1}, 
there exists a basis of sections  
$$
\omega_{i} = 
\sum_{\phi \in \Gamma_{\Delta} / \langle \phi_{i} \rangle} 
\left( \frac{1}{z - \phi(x_{i})} - \frac{1}{z - \phi(x_{-i})} \right) dz \ \ (1 \leq i \leq g) 
$$
of the dualizing sheaf $\omega_{C_{\Delta}/A_{\Delta}}$ on $C_{\Delta}$ 
with coefficients in 
$$
A_{g} \left[ \prod_{k=1}^{g} \frac{1}{(z - x_{k})(z - x_{-k})} \right] [[y_{1},..., y_{g}]]. 
$$
These $\omega_{i}$ are regarded as the universal expression of normalized abelian differentials  since $\omega_{i}|_{x_{\pm h} = a_{\pm h}, y_{h} = q_{h}} = 2 \pi \sqrt{-1} \varphi_{i}$ 
on $R_{\Gamma}$. 
Put 
$$
\omega = \omega_{1} \wedge \cdots \wedge \omega_{g}. 
$$ 

In the case when we consider the Schottky space ${\mathfrak S}_{g}$ 
of degree $g > 1$ as the moduli space of normalized Schottky groups, 
we assume that the above $\phi_{1},..., \phi_{g}$ are {\it normalized} 
by putting $x_{1} = 0$, $x_{-1} = \infty$, namely, 
$$
\phi_{1} = 
\left( \begin{array}{cc} 1 & 0 \\ 0 & y_{1} \end{array} \right) 
\ {\rm mod} \left( B_{\Delta}^{\times} \right), 
$$
and by putting $x_{2} = 1$. 
Then as is shown in \cite[1.1]{I2}, the corresponding generalized Tate curve $C_{\Delta}$ 
is defined over $\tilde{A}_{\Delta} = \tilde{A}_{g} [[y_{1},..., y_{g}]]$, 
where $\tilde{A}_{g}$ is obtained from $A_{g}$ by deleting $x_{-1}$ and 
putting $x_{1} = 0$, $x_{2} = 1$. 
Therefore, one has the associated morphism 
${\rm Spec} \left( \tilde{A}_{\Delta} \right) \rightarrow \overline{\mathcal M}_{g}$. 
\vspace{2ex}

\noindent
{\it 3.3. Explicit formula.} 
For a Schottky group $\Gamma$ such that 
the Hausdorff dimension $\delta_{\Gamma}$ of limit set of $\Gamma$ 
satisfies $\delta_{\Gamma} < 1$, 
the {\it Zograf infinite product} $F_{1}(\Gamma)$ is defined as 
the following absolutely convergent infinite product: 
$$
\prod_{\{ \gamma \}} \prod_{m=0}^{\infty} \left( 1 - q_{\gamma}^{1+m} \right), 
$$
where $\{ \gamma \}$ runs over primitive conjugacy classes in $\Gamma - \{ 1 \}$, 
and $q_{\gamma}$ denotes the multiplier of $\gamma$. 
For an integer $k > 1$ and a marked normalized Schottky group 
$\left( \Gamma; \gamma_{1},..., \gamma_{g} \right)$, 
the {\it McIntyre-Takhtajan infinite product} 
$F_{k}\left( \Gamma; \gamma_{1},..., \gamma_{g} \right)$ is defined as 
the absolutely convergent infinite product 
$$
\left( 1 - q_{\gamma_{1}} \right)^{2} \cdots 
\left( 1 - q_{\gamma_{1}}^{k-1} \right)^{2} 
\left( 1 - q_{\gamma_{2}}^{k-1} \right) 
\prod_{\{ \gamma \}} \prod_{m=0}^{\infty} \left( 1 - q_{\gamma}^{k+m} \right). 
$$

\noindent
{\bf Proposition 3.2.} 
\begin{it} 
Let the notation be as above.  
Then the infinite products 
$$
\prod_{\{ \gamma \}} \prod_{m=0}^{\infty} \left( 1 - q_{\gamma}^{1+m} \right), 
\ \ 
\left( 1 - q_{\gamma_{1}} \right)^{2} \cdots 
\left( 1 - q_{\gamma_{1}}^{k-1} \right)^{2} 
\left( 1 - q_{\gamma_{2}}^{k-1} \right) 
\prod_{\{ \gamma \}} \prod_{m=0}^{\infty} \left( 1 - q_{\gamma}^{k+m} \right) 
$$
have universal expression as invertible elements of $\tilde{A}_{\Delta}$ 
which we denote by $F_{1}, F_{k}$ respectively. 
\end{it} 
\vspace{2ex} 

\noindent
{\it Proof.} 
Let $(\Gamma; \gamma_{1},..., \gamma_{g})$ be a normalized Schottky group, 
and for $i = 1,..., g$, put $\gamma_{-i} = \gamma_{i}^{-1}$. 
Then by Proposition 1.3 of \cite{I2} and its proof, 
if $\gamma \in \Gamma - \{ 1 \}$ has the reduced expression 
$\gamma_{\sigma(1)} \cdots \gamma_{\sigma(l)}$ $(\sigma(i) \in \{ \pm 1,..., \pm g \})$ 
such that $\sigma(1) \neq - \sigma(l)$, 
then its multiplier $q_{\gamma}$ has universal expression as an element 
of $\tilde{A}_{\Delta}$ divisible by $y_{\sigma(1)} \cdots y_{\sigma(l)}$. 
Therefore, the assertion holds. 
\ $\square$ 
\vspace{2ex}

\noindent
{\bf Proposition 3.3.} 
\begin{it} 
There exists a nonzero constant $c(g,k)$ such that 
\end{it} 
$$
c(g,k) \cdot \mu_{g,k} \left( \psi^{(k)} \right) 
= \frac{F_{1}^{d_{k}}}{F_{k}} \omega^{\otimes d_{k}}. 
$$

\noindent
{\it Proof.} 
By Theorem 3.1, $\mu_{g,k}$ gives rise to an isometry 
$$
\lambda_{g,k} \stackrel{\sim}{\rightarrow} 
\lambda_{g,1}^{\otimes d_{k}} \cdot \exp \left( (k - k^{2}) a(g)/2 \right) 
$$
between the metrized tautological line bundles with Quillen metric on ${\mathcal M}_{g}$. 
As is shown by Zograf \cite{Z1, Z2}, 
$F_{1}$ can be extended to a holomorphic function on the Schottky space 
${\mathfrak S}_{g}$ which we denote by the same symbol. 
Let $S_{\rm L}$ denote the classical Liouville action given in Zograf-Takhtajan \cite{ZT}. 
Then by the formula of Zograf \cite{Z1, Z2}, 
$$
\| \omega \|_{\rm Q} \cdot |F_{1}| = 
c_{g} \cdot \exp \left( \frac{S_{\rm L}}{24 \pi} \right), 
$$
and by the formula of McIntyre-Takhtajan \cite{McT}, 
$$
\| \psi^{(k)} \|_{\rm Q} \cdot |F_{k}| = 
c_{g,k} \cdot \exp \left( \frac{S_{\rm L}}{24 \pi} \right)^{d_{k}}, 
$$
where $\| * \|_{\rm Q}$ denotes the Quillen metric,  
$c_{g}$ (resp. $c_{g,k}$) means constants depending only on $g$ (resp. $g, k$). 
Therefore, there exists a holomorphic function $c(g,k)$ on ${\mathfrak S}_{g}$ 
satisfying the above formula such that $\left| c(g,k) \right|$ is a constant function. 
Since ${\mathfrak S}_{g}$ is a connected complex manifold, 
$c(g,k)$ is also a constant function. 
\ $\square$ 
\vspace{2ex}

We consider the case $k = 2$, and give an explicit formula of $\mu_{g,2}$. 
Put 
$$
\left\{ \omega^{(2)}_{1},..., \omega^{(2)}_{3(g-1)} \right\} = 
\left\{ \omega_{l}^{2} \ (1 \leq l \leq g), 
\ \omega_{1} \omega_{l} \ (2 \leq l \leq g),  
\ \omega_{2} \omega_{l} \ (3 \leq l \leq g) \right\}, 
$$
and $\omega^{(2)} = \omega^{(2)}_{1} \wedge \cdots \wedge \omega^{(2)}_{3(g-1)}$.  
Let $\zeta_{i,j}$ $(1 \leq i \leq g, \ 0 \leq j \leq 2)$ be the map from the set 
$\left\{ \phi_{1},..., \phi_{g} \right\}$ of generators of $\Gamma_{\Delta}$ 
into $\tilde{A}_{g}[z]$ defined as 
$$
\zeta_{i,j} \left( \phi_{l} \right) = \delta_{il} (z - x_{i})^{j} 
\ (1 \leq l \leq g). 
$$
Since the coefficients of $\omega^{(2)}_{1},..., \omega^{(2)}_{3(g-1)}$ belong to  $\tilde{A}_{\Delta}$, 
the residue theorem implies that there exists an element 
$\Psi_{g,2} \left( \omega^{(2)}_{m}, \zeta_{i,j} \right)$ of $\tilde{A}_{\Delta}$ such that 
$$
\left. \Psi_{g,2} \left( \omega^{(2)}_{m}, \zeta_{i,j} \right) 
\right|_{x_{\pm h} = a_{\pm h}, y_{h} = q_{h}} = 
\Psi_{g,2} \left( \omega^{(2)}_{m}|_{x_{\pm h} = a_{\pm h}, y_{h} = q_{h}}, 
\zeta_{i,j}|_{x_{h} = a_{h}} \right). 
$$
Therefore, one can define $\det(\Lambda_{g,2}) \in \tilde{A}_{\Delta}$ 
as the determinant of a $3(g-1) \times 3(g-1)$-matrix $\Lambda_{g,2}$ 
consisting of the values of $\Psi_{g,2}$ on  
$$
\left\{ \omega^{(2)}_{1},..., \omega^{(2)}_{3(g-1)} \right\} \times 
\left\{ \zeta_{1,1}, \ \zeta_{2,1}, \ \zeta_{2,2}, \ 
\zeta_{i,0}, \ \zeta_{i,1}, \ \zeta_{i, 2} \ (3 \leq i \leq g) \right\}. 
$$
Note that $\det(\Lambda_{g,2})$ gives (up to a sign) the determinant of the matrix 
consisting of the values of $\Psi_{g,2}$ on 
$$
\left\{ \omega^{(2)}_{1},..., \omega^{(2)}_{3(g-1)} \right\} \times 
\left\{ \xi_{1,1}, \ \xi_{2,1}, \ \xi_{2,2}, \ \xi_{i,0}, \ \xi_{i,1}, \ \xi_{i, 2} \ (3 \leq i \leq g) \right\}, 
$$ 
and hence it gives the universal expression of the values of $\det (\Psi_{g,2})$ 
on the products of normalized abelian differentials. 
\vspace{2ex}

\noindent
{\bf Theorem 3.4.}  
\begin{it} 
Let the notation be as above. 
Then 
\end{it} 
$$
\mu_{g,2} \left( \omega^{(2)} \right) = 
\pm \det(\Lambda_{g,2}) \frac{F_{1}^{13}}{F_{2}} \omega^{\otimes 13}. 
$$

\noindent
{\it Proof.} 
Since $\left\{ \omega^{(2)}_{i} \right\}$ is a basis of the space of regular $2$-forms 
on genus $2$ or non-hyperelliptic curves, 
$\det(\Lambda_{g,2})$ is non-zero and $\omega^{(2)}/ \det(\Lambda_{g,2})$ gives 
the exterior product $\psi^{(2)}$ of the normalized $2$-differentials 
on ${\mathfrak S}_{g}$. 
Then by Theorem 3.1 and Propositions 3.2, 3.3,  
$\mu_{g,2} \left( \omega^{(2)} \right)$, 
$\det(\Lambda_{g,2}) F_{1}^{13} F_{2}^{-1} \omega^{\otimes 13}$ are 
represented by elements of $\tilde{A}_{\Delta}$, 
and are equal to each other up to a non-zero constant. 
Hence this constant is a rational number. 
Furthermore, 
Proposition 3.2 implies that $F_{1}^{13} F_{2}^{-1} \omega^{\otimes 13}$ is primitive, 
namely is not congruent to $0$ modulo any prime number. 
In the following, 
we will prove that $\det(\Lambda_{g,2})$ is primitive in $\tilde{A}_{\Delta}$ 
by calculating its leading term. 
Then $\omega^{(2)}_{1},..., \omega^{(2)}_{3(g-1)}$ are linearly independent 
over the quotient field of $\tilde{A}_{\Delta} \widehat{\otimes} K$ 
for a field $K$ of any characteristic, 
and hence $\mu_{g,2} \left( \omega^{(2)} \right)$ is primitive. 
Therefore, the assertion follows from that 
$\det(\Lambda_{g,2})$ is primitive in $\tilde{A}_{\Delta}$. 

If $\psi (dz)^{2}$ is a $2$-differential 
on a Schottky uniformized Riemann surface  $R_{\Gamma}$, 
then $\Psi_{g,2} \left( \psi, \zeta_{i,j} \right)$ is the sum of the residues of 
$\psi (z - x_{i})^{j} dz$ in the interior of $C_{i}$. 
By \cite[Proposition 3.2]{I1},  
$$
\omega_{i} = 
\left( \frac{1}{z - x_{i}} - \frac{1}{z - x_{-i}} + 
\sum_{\phi \in \Gamma_{\Delta} / \langle \phi_{i} \rangle - \{ 1 \}} 
\frac{\phi(x_{i}) - \phi(x_{-i})}{(z - \phi(x_{i}))(z - \phi(x_{-i}))} \right) dz, 
$$
and $\phi(x_{i}) - \phi(x_{-i}) \in \prod_{j=1}^{l} y_{|\sigma(j)|} \cdot \tilde{A}_{\Delta}$ 
if $\phi$ has the reduced expression $\phi_{\sigma(1)} \cdots \phi_{\sigma(l)}$, 
where $\sigma(j) \in \{ \pm 1,..., \pm g \}$, 
$\phi_{-\sigma(j)} = \phi_{\sigma(j)}^{-1}$ and $\sigma(l) \neq \pm i$. 
Denote by $I_{\Delta}$ the ideal of $\tilde{A}_{\Delta}$ generated by $y_{1},..., y_{g}$.  
Then we have  
$$
\Psi_{g,2} \left( \omega_{i}^{2}, \zeta_{i,1} \right) \equiv 
{\rm Res}_{x_{i}} \frac{z - x_{i}}{(z - x_{i})^{2}} = 
1 \ {\rm mod} \left( I_{\Delta} \right), 
$$
and for $l \neq i$, 
\begin{eqnarray*}
\Psi_{g,2} \left( \omega_{i} \omega_{l}, \zeta_{i,0} \right) 
& \equiv & 
{\rm Res}_{x_{i}} 
\left( \frac{1}{z - x_{i}} - \frac{1}{z - x_{-i}} \right) 
\left( \frac{1}{z - x_{l}} - \frac{1}{z - x_{-l}} \right) 
\\
& = & 
\frac{x_{l} - x_{-l}}{(x_{i} - x_{l})(x_{i} - x_{-l})} \ 
{\rm mod} \left( I_{\Delta} \right). 
\end{eqnarray*}
If $l \neq i$, then 
$$
\phi_{i}(x_{l})^{\pm 1} \equiv 
x_{\pm i} + \frac{(x_{l} - x_{\pm i})(x_{\pm i} - x_{\mp i})}{(x_{l} - x_{\mp i})} y_{i} 
\ {\rm mod} \left( I_{\Delta}^{2} \right), 
$$ 
and hence 
\begin{eqnarray*} 
& & \Psi_{g,2} \left( \omega_{i} \omega_{l}, \zeta_{i,2} \right)  
\\
& = & 
{\rm Res}_{\phi_{i}(x_{l})} 
\left( \frac{1}{z - x_{i}} - \frac{1}{z - x_{-i}} \right) 
\left( \frac{1}{z - \phi_{i}(x_{l})} - \frac{1}{z - \phi_{i}(x_{-l})} \right) 
(z - x_{i})^{2} 
\\
& & + \ 
{\rm Res}_{\phi_{i}(x_{-l})} 
\left( \frac{1}{z - x_{i}} - \frac{1}{z - x_{-i}} \right) 
\left( \frac{1}{z - \phi_{i}(x_{l})} - \frac{1}{z - \phi_{i}(x_{-l})} \right) 
(z - x_{i})^{2} 
\\ 
& & + \ \cdots 
\\
& = & 
{\rm Res}_{\phi_{i}(x_{l})} \frac{1}{z - \phi_{i}(x_{l})}
\left( z - x_{i} - \frac{(z - x_{i})^{2}}{z - x_{-i}} \right) 
\\ 
& & + \ 
{\rm Res}_{\phi_{i}(x_{-l})} \frac{1}{z - \phi_{i}(x_{-l})}
\left( z - x_{i} - \frac{(z - x_{i})^{2}}{z - x_{-i}} \right) 
+ \cdots 
\\
& = & 
\phi_{i}(x_{l}) - \phi_{i}(x_{-l}) + \cdots 
\\
& = & 
\frac{(x_{l} - x_{-l}) (x_{i} - x_{-i})^{2}}
{(x_{l} - x_{-i})(x_{-l} - x_{-i})} y_{i} + \cdots. 
\end{eqnarray*}
Furthermore, for $l,m \neq i$, 
\begin{eqnarray*}
& & 
\Psi_{g,2} \left( \omega_{l} \omega_{m}, \zeta_{i,2} \right)  
\\ 
& = & 
{\rm Res}_{\phi_{i}(x_{l})} \frac{1}{z - \phi_{i}(x_{l})} 
\left( \frac{1}{z - x_{m}} - \frac{1}{z - x_{-m}} \right) (z - x_{i})^{2} 
\\
& & - \ {\rm Res}_{\phi_{i}(x_{-l})} \frac{1}{z - \phi_{i}(x_{-l})} 
\left( \frac{1}{z - x_{m}} - \frac{1}{z - x_{-m}} \right) (z - x_{i})^{2} 
\\
& & + \ {\rm Res}_{\phi_{i}(x_{m})} 
\left( \frac{1}{z - x_{l}} - \frac{1}{z - x_{-l}} \right) 
\frac{(z - x_{i})^{2}}{z - \phi_{i}(x_{m})} 
\\
& & - \ {\rm Res}_{\phi_{i}(x_{-m})} 
\left( \frac{1}{z - x_{l}} - \frac{1}{z - x_{-l}} \right) 
\frac{(z - x_{i})^{2}}{z - \phi_{i}(x_{-m})} + \cdots 
\\
& = & 
\left( \frac{1}{x_{i} - x_{m}} - \frac{1}{x_{i} - x_{-m}} \right) 
\left\{ \left( \phi_{i}(x_{l}) - x_{i} \right)^{2} - 
\left( \phi_{i}(x_{-l}) - x_{i} \right)^{2} \right\} 
\\
& & + \ 
\left( \frac{1}{x_{i} - x_{l}} - \frac{1}{x_{i} - x_{-l}} \right) 
\left\{ \left( \phi_{i}(x_{m}) - x_{i} \right)^{2} - 
\left( \phi_{i}(x_{-m}) - x_{i} \right)^{2} \right\} + \cdots 
\\
& = & 
\frac{(x_{i} - x_{-i})^{2} (x_{m} - x_{-m})}{(x_{i} - x_{m})(x_{i} - x_{-m})} 
\left\{ \left( \frac{x_{l} - x_{i}}{x_{l} - x_{-i}} \right)^{2} - 
\left( \frac{x_{-l} - x_{i}}{x_{-l} - x_{-i}} \right)^{2} \right\} y_{i}^{2} 
\\
& & + \ 
\frac{(x_{i} - x_{-i})^{2} (x_{l} - x_{-l})}{(x_{i} - x_{l})(x_{i} - x_{-l})} 
\left\{ \left( \frac{x_{m} - x_{i}}{x_{m} - x_{-i}} \right)^{2} - 
\left( \frac{x_{-m} - x_{i}}{x_{-m} - x_{-i}} \right)^{2} \right\} y_{i}^{2} + 
\cdots. 
\end{eqnarray*}
Therefore, we have 
$$
\left\{ \begin{array}{ll} 
\Psi_{g,2} \left( \omega_{1} \omega_{l}, \zeta_{i,0} \right) \equiv 
{\displaystyle \frac{\delta_{il}(x_{1} - x_{-1})}
{(x_{i} - x_{1})(x_{i} - x_{-1})}} \ {\rm mod} \left( I_{\Delta} \right) 
& (2 \leq l \leq g), 
\\ 
\Psi_{g,2} \left( \omega_{2} \omega_{l}, \zeta_{i,0} \right) \equiv 
{\displaystyle \frac{\delta_{il}(x_{2} - x_{-2})}
{(x_{i} - x_{2})(x_{i} - x_{-2})}} \ {\rm mod} \left( I_{\Delta} \right) 
& (3 \leq l \leq g) 
\end{array} \right. 
$$
for $3 \leq i \leq g$, 
$$
\left\{ \begin{array}{ll} 
\Psi_{g,2} \left( \omega_{l}^{2}, \zeta_{i,1} \right) \equiv 
\delta_{il} \ {\rm mod} \left( I_{\Delta} \right) 
& (1 \leq l \leq g), 
\\
\Psi_{g,2} \left( \omega_{1} \omega_{l}, \zeta_{i,1} \right) \equiv 
0 \ {\rm mod} \left( I_{\Delta} \right) 
& (2 \leq l \leq g), 
\\
\Psi_{g,2} \left( \omega_{2} \omega_{l}, \zeta_{i,1} \right) \equiv 
0 \ {\rm mod} \left( I_{\Delta} \right) 
& (3 \leq l \leq g) 
\end{array} \right. 
$$
for $1 \leq i \leq g$, 
and 
$$
\left\{ \begin{array}{ll} 
\Psi_{g,2} \left( \omega_{1} \omega_{l}, \zeta_{i,2} \right) \equiv 
{\displaystyle 
\frac{\delta_{il}(x_{1} - x_{-1})(x_{i} - x_{-i})^{2}}
{(x_{-i} - x_{1})(x_{-i} - x_{-1})} y_{i}} 
\ {\rm mod} \left( I_{\Delta}^{2} \right) 
& (2 \leq l \leq g), 
\\
\Psi_{g,2} \left( \omega_{2} \omega_{l}, \zeta_{2,2} \right) \equiv 
{\displaystyle 
\frac{(x_{l} - x_{-l})(x_{2} - x_{-2})^{2}}
{(x_{l} - x_{-2})(x_{-l} - x_{-2})} y_{2}} 
\ {\rm mod} \left( I_{\Delta}^{2} \right), 
& 
\\
\Psi_{g,2} \left( \omega_{2} \omega_{l}, \zeta_{i,2} \right) \equiv 
{\displaystyle 
\frac{\delta_{il}(x_{2} - x_{-2})(x_{i} - x_{-i})^{2}}
{(x_{-i} - x_{2})(x_{-i} - x_{-2})} y_{i}} 
\ {\rm mod} \left( I_{\Delta}^{2} \right) 
& (i \neq 2, \ 3 \leq l \leq g) 
\end{array} \right. 
$$
for $2 \leq i \leq g$. 
Hence the leading term of $\det(\Lambda_{g,2}) \in \tilde{A}_{\Delta}$ is 
${\displaystyle \pm \prod_{i=2}^{g} \tau_{i}}$, 
where 
$$
\tau_{2} \equiv 
\Psi_{g,2} \left( \omega_{2} \omega_{1}, \zeta_{2,2} \right) \ 
{\rm mod} \left( I_{\Delta} \right) = 
\frac{(x_{1} - x_{-1})(x_{2} - x_{-2})^{2}}{(x_{1} - x_{-2})(x_{-1} - x_{-2})} y_{2}, 
$$
and for $3 \leq i \leq g$, 
\begin{eqnarray*}
\tau_{i} 
& \equiv & 
\Psi_{g,2} \left( \omega_{i} \omega_{2}, \zeta_{i,0} \right) 
\Psi_{g,2} \left( \omega_{i} \omega_{1}, \zeta_{i,2} \right) - 
\Psi_{g,2} \left( \omega_{i} \omega_{1}, \zeta_{i,0} \right) 
\Psi_{g,2} \left( \omega_{i} \omega_{2}, \zeta_{i,2} \right)
\ {\rm mod} \left( I_{\Delta}^{2} \right) 
\\
& = & 
\left( 
\frac{1}{(x_{-i} - x_{1})(x_{-i} - x_{-1})} - 
\frac{(x_{i} - x_{2})(x_{i} - x_{-2})}
{(x_{-i} - x_{2})(x_{-i} - x_{-2})(x_{i} - x_{1})(x_{i} - x_{-1})} 
\right) 
\\
& & \times \ 
\frac{(x_{1} - x_{-1})(x_{2} - x_{-2})(x_{i} - x_{-i})^{2}}{(x_{i} - x_{2})(x_{i} - x_{-2})} y_{i}. 
\end{eqnarray*}
Therefore, this product is primitive in $\tilde{A}_{\Delta}$, 
and hence the assertion holds. 
\ $\square$

\end{document}